\documentclass[aps,prl,twocolumn,floatfix,english,superscriptaddress,showpacs,10pt]{revtex4-2} 
\usepackage{amsmath,bm}
\usepackage{mathtools}
\usepackage{commath}
\usepackage{amssymb}
\usepackage{graphicx}
\usepackage{babel}
\usepackage{cases}
\usepackage[colorlinks=true, citecolor=blue, anchorcolor=green]{hyperref}
\hypersetup{linkcolor=red}
\usepackage{natbib}
\usepackage{floatrow}
\usepackage{dblfloatfix}
\usepackage{xcolor}
\usepackage{braket}
\usepackage{lipsum}

\makeatletter

\newcommand{\Rmnum}[1]{\expandafter\@slowromancap\romannumeral #1@}

\makeatother

\begin{document}

\title{Surface Ferron Excitations in Ferroelectrics and Their Directional Routing  }
\author{Xi-Han Zhou}
\affiliation{School of Physics, Huazhong University of Science and Technology, Wuhan 430074, China}
\author{Chengyuan Cai}
\affiliation{School of Physics, Huazhong University of Science and Technology, Wuhan 430074, China}
\author{Ping Tang}
\affiliation{WPI-AIMR, Tohoku University, 2-1-1 Katahira, 980-8577 Sendai, Japan}
\author{R. L. Rodr\'iguez-Su\'arez}
\affiliation{Facultad de F\'isica, Pontificia Universidad Cat\'olica de Chile, Casilla 306, Santiago, Chile}
\author{S. M. Rezende}
\affiliation{Departamento de F\'isica, Universidade Federal de Pernambuco, 50670-901, Recife, Pernambuco, Brazil}
\author{G. E. W. Bauer}
\affiliation{WPI-AIMR and Institute for Materials Research and CSRN, Tohoku University, Sendai 980-8577, Japan}
 \affiliation{Kavli Institute for Theoretical Sciences, University of the Chinese Academy of Sciences, Beijing 10090, China}
\author{Tao Yu}
\email{taoyuphy@hust.edu.cn}
\affiliation{School of Physics, Huazhong University of Science and Technology, Wuhan 430074, China}

\date{\today}

\begin{abstract}
The duality between electric and magnetic dipoles inspires recent comparisons  between ferronics and magnonics. Here we predict surface polarization waves or ``ferrons" in ferroelectric insulators, taking the long-range dipolar interaction into account. We predict properties that are strikingly different from the magnetic  counterpart, \textit{i.e.} the surface ``Damon-Eshbach" magnons in ferromagnets. The dipolar interaction pushes the ferron branch with locked circular polarization and momentum to the ionic plasma frequency. The low-frequency modes are on the other hand in-plane polarized normal to their wave vectors. The strong anisotropy of the lower branch renders directional emissions of electric polarization and chiral near fields when activated by a focused laser beam, allowing optical routing in ferroelectric devices.
\end{abstract}

\maketitle

\textit{Introduction.}---Electric and magnetic dipoles are said to be dual since the associated electric ${\bf E}$ and magnetic ${\bf H}$ fields have identical forms \cite{Duality,Jackson}. The elementary excitations of ordered dipoles in both magnets and ferroelectrics can be classified as massless Goldstone modes that restore the broken symmetry and Higgs modes, i.e. scalar excitation of the modulus of the order parameter. Magnons are transverse spin excitation the classify as Goldstone modes in the absence of applied or anisotropy fields. The polarization waves of the ferroelectric order, or ``ferrons" \cite{ferron_definition}, can have either character, depending on the microscopic mechanism that triggers the ferroelectric phase transition.

Magnons in magnetic insulators favor low-dissipation \cite{roadmap,PR_insulator} and novel chiral \cite{PR_chirality} transport for the spin, momentum, and heat. Among them, Damon-Eshbach (DE) modes in finite-size magnets, governed by the long-range  magnetic dipolar interaction, are surface modes holding chirality ruled by the cross product of magnetization, propagation direction, and surface normal \cite{DE,Walker_sphere}. Being unidirectional normal to the magnetization, they can propagate against temperature gradients \cite{heat_conveyer_1,heat_conveyer_2,heat_conveyer_3,heat_conveyer_4}, be backscattering immune to the surface disorder \cite{disorder_1,disorder_2}, and interact chirally with light \cite{Usami,Sanchar}. Such modes are recently demonstrated to be topological, even
though they do not lie in a gap \cite{Kei}.
Surface modes of ferroelectrics  that emerge in an exchange-only transverse Ising model  \cite{surface_ferron}  do not take into account electric dipolar interactions  and have not yet been verified by experiments. Recent theoretical proposals predict thermal transport of electric polarization by bulk ferrons in nanostructures \cite{ferron_1,ferron_2,ferron_3,ferron_4}, \textcolor{blue}{with initial possible experimental evidence \cite{first_experiment}}.  Magnetism and ferroelectricity respectively break time-reversal and inversion symmetry \cite{gibbs}. 

\begin{figure}[t]
	 \centering
	 \includegraphics[width=0.9\linewidth]{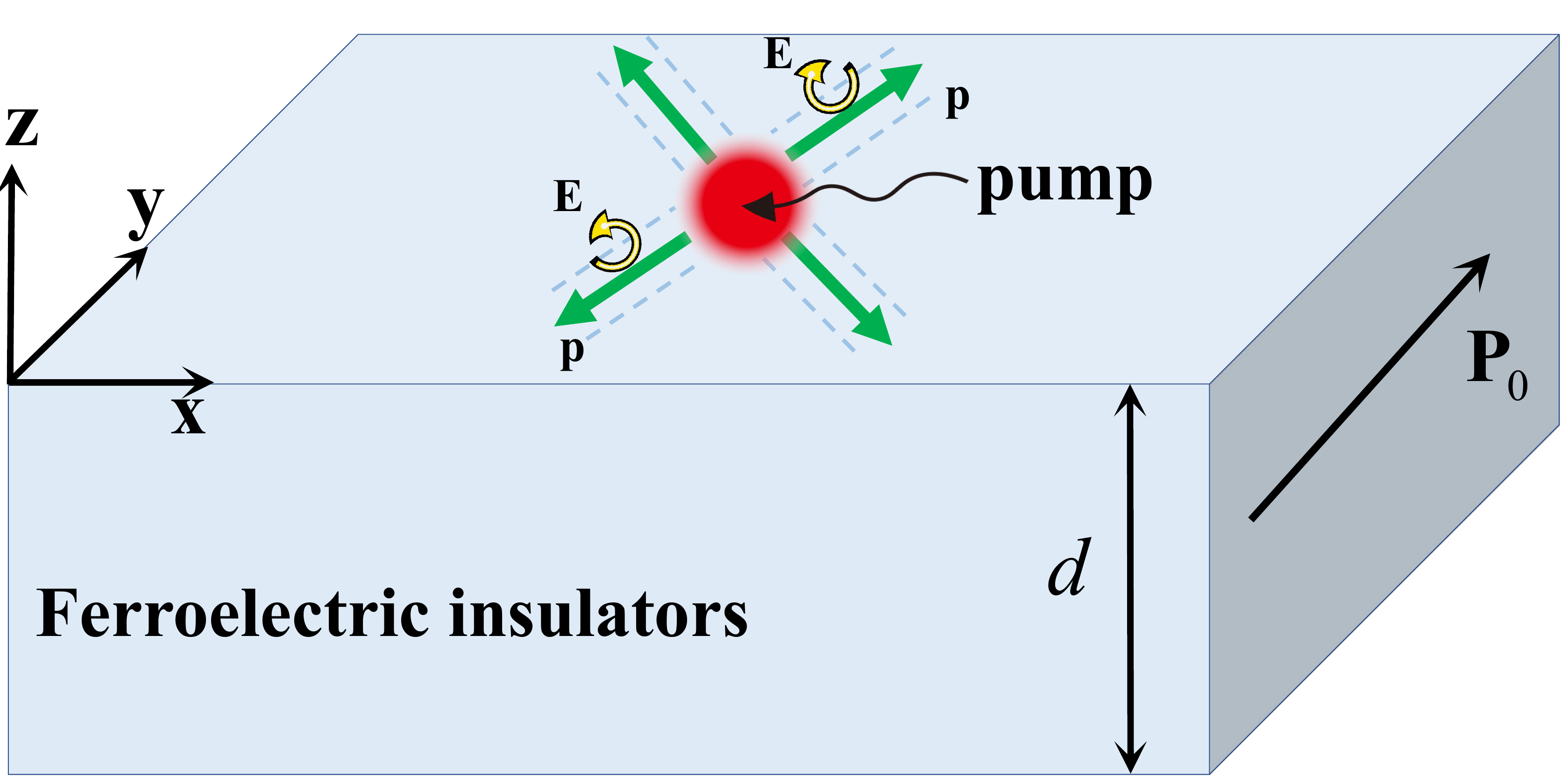}
	 \caption{Surface ferrons excited by a laser spot (red dot) in a ferroelectric insulator of thickness $d$ with polarization ${\bf P}_0$ poled along the $\hat{\bf y}$-direction. The green arrows illustrate the predicted emission of directional coherent ferron beams with four-fold symmetry, and the yellow arrows illustrate the associated electric stray fields with momentum-locked polarization.}
	 \label{fig:model}
	 \end{figure}

In this Letter, we predict surface ferrons in ferroelectrics that exist by the long-range dipolar interaction. They generate strong chiral stray electric fields in the THz regime, as sketched in Fig.~\ref{fig:model}.
We find two branches of surface ferrons, among which one branch is pushed to the surface ionic plasma frequency $\sim \Omega_p/\sqrt{2}$ by the crucial dipolar interaction, but the other branch at much lower frequency $\sim \Omega_p/10\sim 10$~THz remains. The polarization of the lower mode is linear and lies in the surface plane transverse to the wave vector. In contrast to the  DE magnons in ferromagnets, the surface ferrons are therefore not chiral. Nevertheless, their stray electric fields are circularly polarized with the precession axis normal and locked to their momenta. Such chirality has been found as well in conventional surface polaritons and Rayleigh surface acoustic waves  \cite{PR_chirality,Weyl,chiral_optics,Nori_review,Matsuo_review,SAW,SOC_light}. We predict that the dispersion relations of surface ferrons are highly anisotropic leading to directional emission from a focused excitation as illustrated by Fig.~\ref{fig:model}. These results imply useful functionalities for broadband optical micro-scale routing in electrically insulating rather than metallic devices \cite{plasmon_1,plasmon_2,plasmon_3,plasmon_4,nanorouting_1,nanorouting_2}.

\textit{Model and formalism}.---We consider a slab of uniaxial ferroelectric insulator of thickness $ d $ in vacuum that is extended in the $x$-$y$ plane and normal to the $\hat{\bf z}$-axis (Fig.~\ref{fig:model}). The spontaneous polarization $ {\bf P}_0 $ at equilibrium is poled along the $\hat{ \bf y }$-direction and emits no stray fields.  In electrical insulators, the bare Coulomb interaction between fluctuations of the electric polarizations ${\bf p}({\bf r})$ is strong and long-ranged, which renders the problem nontrivial. Assuming small fluctuations $|{\bf p}|\ll|{\bf P}_0|$ we find the collective modes by self-consistently solving the Landau-Khalatnikov-Tani (LKT) equation of motion of the electric polarization \cite{LKTequation,LKT2,LKT3,LKT4,GL1,GL2,GL3,GL4,PbTiO3,LiNbO3,PbTiO3-1,PRB_dipolar} coupled to the Maxwell equation for the electric field ${\bf E}^{(d)}$ emitted by ${\bf p}$.
	 
We start from the free energy $F=\int d{\bf r}{\mathcal F}({\bf r})$ below the Curie temperature $T_c$ of a uniaxial ferroelectric insulator biased by an external (DC or AC) electric field $ {\bf E}^{(e)} $, with the energy density as a function of the electric polarization ${\bf P}={\bf P}_0+{\bf p}$ \cite{gibbs,PRB_dipolar,LiNbO3,LiTaO_3}
		\begin{align}
			{\mathcal F}({\bf r})&=({\alpha}/{2}) P_y^{2}+({\beta}/{4}) P_y^{4}+({\lambda}/{2}) (P_x^2+P_z^2)\nonumber\\
		&-({1}/{2}){\bf E}^{(d)}\cdot{\bf P} -{\bf E}^{(e)} \cdot {\bf P}.
		\label{free_energy}
		\end{align}	
   Our results for $T \ll T_c$ do not depend importantly on Landau parameters $ \alpha<0$, $\beta>0$, and $\lambda >0$. We are interested in the surface ferrons and their difference from conventional phonons in dielectrics \cite{PbTiO3-1,surface_phonon_polariton_0,surface_phonon_polariton_1,surface_phonon_polariton_2,LKT3}. The electric polarization is an independent thermodynamic variable with independent Goldstone and Higgs excitations different from that of conventional phonons, noting that in displacive ferroelectrics the ferron equations of motion are closely related to lattice dynamics, however: The electric polarization $\sum_i{Q}_i{\bf r}_i$ by atoms positioned at ${\bf r}_i$ with effective charges $Q_i$ is governed by the ionic motion driven by effective electric fields \cite{LKT3,PRB_Letter} and elastic forces \cite{surface_phonon_polariton_0,surface_phonon_polariton_1,surface_phonon_polariton_2}. \textcolor{blue}{ In displacive ferroelectrics, phonon dynamics governs the Landau parameters. On the other hand, purely electronic ferroelectricity can be described by a Landau functional as well \cite{electronic_ferroelectricity}. }

   Here we focus on the LKT equation of motion \cite{LKTequation,LKT2,LKT3,LKT4} 
	\begin{align}
		m_{p} \partial^{2}{\bf P}/ \partial t^{2}+\gamma {\partial {\bf P}}/{\partial t}=-{\delta F}/{\delta{\bf P}},
		\label{LKTeq}
	\end{align}
which is interpreted as a damped Newton's equation for an effective mass $ m_p=1/(\varepsilon_0 \Omega_p^2) $, where $\varepsilon_0$ is the vacuum dielectric constant, ${\Omega_p}$ is the \textit{ionic} plasma frequency, and $\gamma $ is a phenomenological viscous damping constant \cite{LKT3}. 
 ${\bf P}$ is a source for the electric field that obeys the wave equation \cite{Jackson}
$\nabla^2{\bf E}^{(d)}-({1}/{c^2})\partial^2_t{\bf E}^{(d)}=\mu_0\partial^2_t{\bf P}-({1}/{\varepsilon_0})\nabla(\nabla\cdot{\bf P})$,
contributed by both the electromagnetic radiation via the shift current density $\dot{\bf P}$ and Coulomb's law for the bound charge accumulation $-\nabla\cdot {\bf P}$, where $\mu_0$ is the vacuum permeability.
For a monochromatic fluctuation ${\bf P}({\bf r},t) \propto e^{-i\omega t}$ and assuming that 
its wavelength $\lambda$ is much \textit{smaller} than that of light, $k\gg \omega/c$ and $\lambda \ll 2\pi c/\omega\equiv \lambda_c$, they do not couple strongly because of the wave vector mismatch. At the frequency band of our interest $\omega\sim 10$~THz, $\lambda_c\sim 190~\mu$m is a long wavelength. The retardation due to the finite light velocity $c$ is negligible and we may treat the electric dipolar field in the static limit \cite{Jackson}
	\begin{equation}
		{\bf E}^{(d)}({\bf r},t)=\dfrac{1}{4\pi \varepsilon_0}\nabla\int \dfrac{\nabla'\cdot{\bf P}({\bf r}',t)}{|{\bf r}-{\bf r}'|}d{\bf r}'.
		\label{dipolar_field}
	\end{equation}
At equilibrium, a constant electric polarization $ {\bf P}_0 =\left\lbrace 0,P_0,0\right\rbrace$ minimizes the free energy $ \delta F/\delta {\bf P}=0 $ when saturated by the static external field $E^{(e)}_y\hat{\bf y}$, and according to $\alpha P_0+\beta P_0^3=E^{(e)}_y$, $ P_{0}^{2}\rightarrow -\alpha/\beta $ when $ E_y^{(e)}\rightarrow 0$.

 In the linear regime with $|{\bf p}|\ll |{\bf P}_0|$ and assuming a small external electric field ${\bf E}^{(e)}\rightarrow 0$, the equation of motion for the fluctuations reads
	\begin{align}
	&(1/\Omega_p)^2\partial_t^{2}p_{x,z}+K_{\perp} p_{x,z} =\varepsilon_0E_{x,z}^{(d)},\nonumber\\
	&(1/\Omega_p)^2 \partial_t^{2}p_y+K_{\parallel} p_y=\varepsilon_0E_y^{(d)},
\label{EOM_fluctuation}
\end{align}	
where $K_{\perp}=\varepsilon_0\lambda>0$ and $K_{\parallel}=\varepsilon_0(\alpha+3 \beta P_0^2 )\rightarrow -2\varepsilon_0\alpha>0$ are dimensionless ``stiffness" constants. The two branches of the eigenmodes describe longitudinal (Higgs-like) and transverse (Goldstone-like) fluctuations. Reference~\cite{ferron_3} addressed only longitudinal modes, disregarding the dipolar interaction that couples them with the transverse ones. The Coulomb interaction gaps the Goldstone mode at zero wave number to the ionic plasma frequency $\Omega_p$ as derived below [see also the Supplemental Material (SM) \cite{supplement}], analogous to the magnon gap induced by magnetic anisotropies and the  Goldstone mode in superconductors that is pushed to the electron plasma frequency by charge fluctuations  \cite{Anderson,Nambu}.

Equations~(\ref{dipolar_field}) and (\ref{EOM_fluctuation}) describe the interplay of the electric polarization and its dipolar electric field. We expand the solution into plane waves
	\begin{align}
	 {\bf p}({\bf r},t)=\left({\bf a}e^{i k_z z}+{\bf b}e^{-i k_z z}\right)e^{i ({\pmb \kappa}\cdot{\pmb \rho}-\omega t)},
	 \label{plane_wave}
	\end{align}
where the in-plane coordinate $ {\pmb \rho}=x \hat{\mathbf{x}} + y \hat{\mathbf{y}} $ and momentum ${\pmb \kappa}=k_x\hat{\bf x}+k_y\hat{\bf y}$, which generates the stray field \cite{Weyl,PR_chirality}
	\begin{equation}
	{\bf E}^{(d)}=\frac{e^{i ({\pmb \kappa}\cdot{\pmb \rho}-\omega t)}}{	\varepsilon_0 } \int_{-d}^{0}dz'\mathcal{G}(z-z')\left( \begin{array}{c}
			a_xe^{ik_z z'}+b_xe^{-i k_z z'} \\
	a_ye^{ik_z z'}+b_ye^{-i k_z z'}	 \\
		a_ze^{ik_z z'}+b_ze^{-i k_z z'}
		\end{array}\right), 
		\label{integral_G}
	\end{equation}
where $\mathcal{G}(z-z')$ is the Green-function tensor. Here ${k_y}E_x^{(d)}={k_x}E_y^{(d)}$ implies locking of the circular polarization and wave vector (see SM \cite{supplement} for details). 

\textit{Surface ferrons.}---The polarization waves interact with their self-fields caused by the bound surface and bulk electric charges that must be calculated self-consistently. Substituting Eq.~(\ref{integral_G}) into (\ref{EOM_fluctuation}), we find two coupled characteristic equations (refer to the SM \cite{supplement} for details of the derivation) that can solve $\omega$ and $k_z$ in terms of $k_x$ and $k_y$. The first characteristic equation 
\begin{equation}
	c_1 c_2k^2+c_1 k_y^2+c_2 (k_x^2+k_z^2)=0
	\label{characteristic_1}
\end{equation}
establishes the relation between the frequency and momentum,
where $ c_1\equiv -(\omega/\Omega_p)^2+K_{\perp}$ and $c_2\equiv -(\omega/\Omega_p)^2+K_{\parallel}$ are dimensionless. For bulk excitation, $k_z$ is real and Eq.~(\ref{characteristic_1}) leads to two ferron branches with band gaps at  $\omega\sim \Omega_p$ and $\omega\ll \Omega_p$, respectively  \cite{supplement}.  The high-frequency branch is longitudinal with ${\bf p}\parallel {\bf k}$ \cite{supplement}, with its frequencies pushed up to the bulk ionic plasma frequency $\Omega_p$ by the Anderson-Higgs mechanism \cite{Anderson,Nambu}. The low-frequency branch is a mix of both  longitudinal \cite{ferron_3} and transverse modes. A complex $k_z=\eta_1+i\eta_2$ that also obeys the second characteristic equation \cite{supplement}
 \begin{align}
 \nonumber 
 		&\left[
 	(c_1+1)k_z\sin({k_zd}/{2})-c_1\kappa\cos(k_zd/2)
 	 \right] \\
 		&\times\left[
 	(c_1+1)k_z\cos(k_zd/2)+ c_1\kappa\sin(k_zd/2)
 	 \right]=0,
 	 \label{characteristic_2}
 	\end{align}
implying the existence of evanescent modes at the surfaces. We solve Eqs.~(\ref{characteristic_1}) and (\ref{characteristic_2}) numerically for ferroelectric films of arbitrary thickness. 
 
For thick slabs $\coth{(d\eta_2/2})\rightarrow 1$ and $\tanh({d\eta_2/2})\rightarrow1$, so according to Eq.~(\ref{characteristic_2}), $\eta_1=0 $ and $\eta_2=-{c_1\kappa}/{(1+c_1)}$ \cite{supplement}.
 Substituting $ k_z=\pm i\eta_2 $ into Eq.~(\ref{characteristic_1})  reduces the coupled characteristic Eqs.~(\ref{characteristic_1}) and (\ref{characteristic_2}) to  
	\begin{align}
	\nonumber
		&k_z=\pm i{c_1\kappa}/(1+c_1),\\
		& k_x^2c_2(1+2c_1)+k_y^2\left[c_1c_2+c_1(1+c_1)\right]  =0.
		\label{simplified}
	\end{align}
In most ferroelectrics, the elastic constants and their difference $\delta=K_\parallel-K_\perp$ are much smaller than unity at room temperature. Solving Eq.~(\ref{simplified}) to leading order in $\delta$ leads to surface modes
\begin{align}
	 \omega_{\pmb \kappa}^+&=\Omega_p\sqrt{(1+\delta\cos^2{\theta_{\pmb \kappa}})/2+K_\perp}, \nonumber\\
	 {\omega_{\pmb \kappa}^-}&={\Omega_p}\sqrt{\delta\sin^2{\theta_{\pmb \kappa}}+K_\perp},
	 \label{frequencies}
\end{align}
that are  gaped and depend on the angle $ \theta_{\pmb \kappa}$ between propagation direction and equilibrium polarization. 

\begin{figure}[htp]
	\centering
 \hspace{-0.145cm}\includegraphics[width=0.495\linewidth]{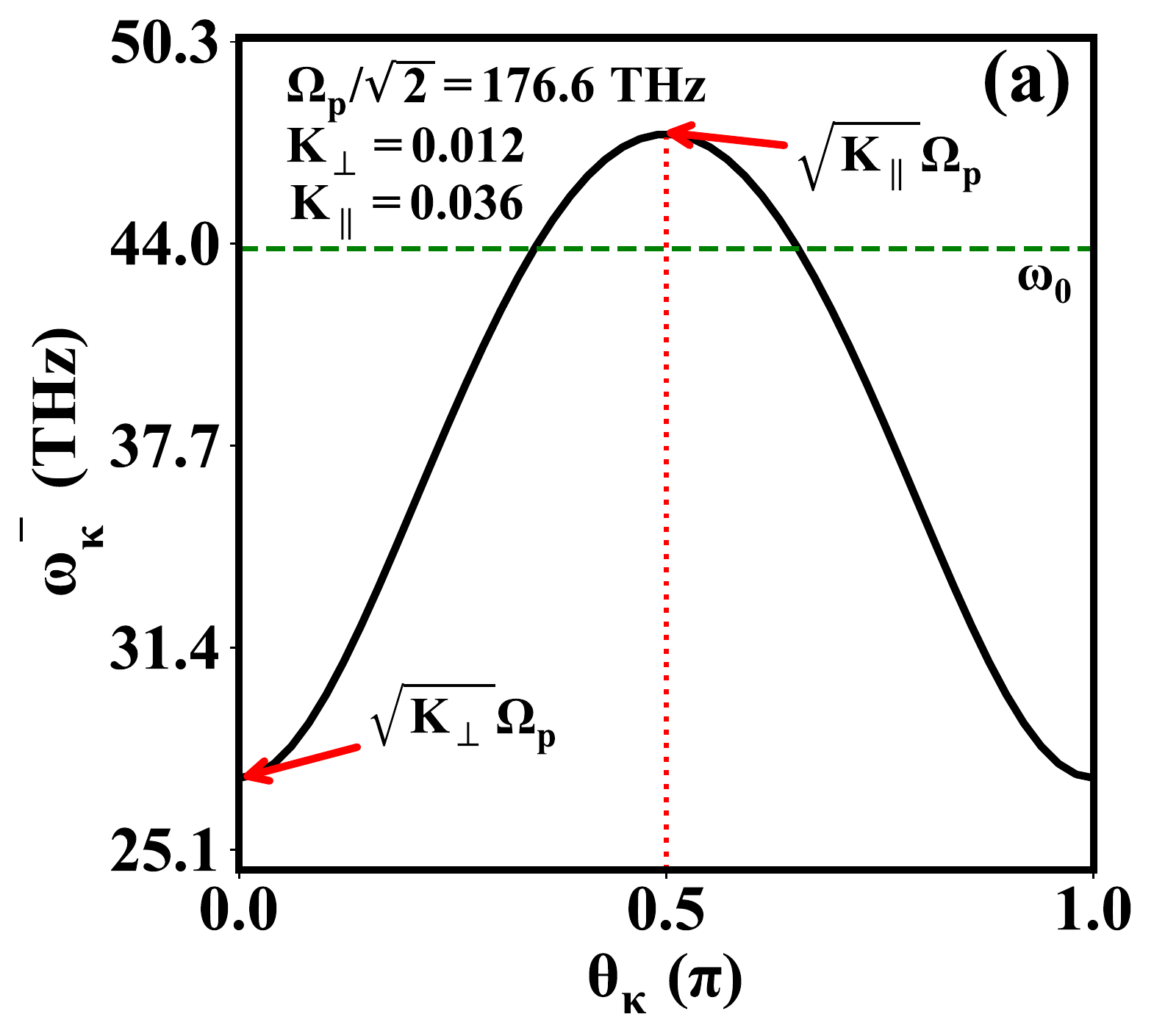}
\hspace{-0.26cm}\includegraphics[width=0.515\linewidth]{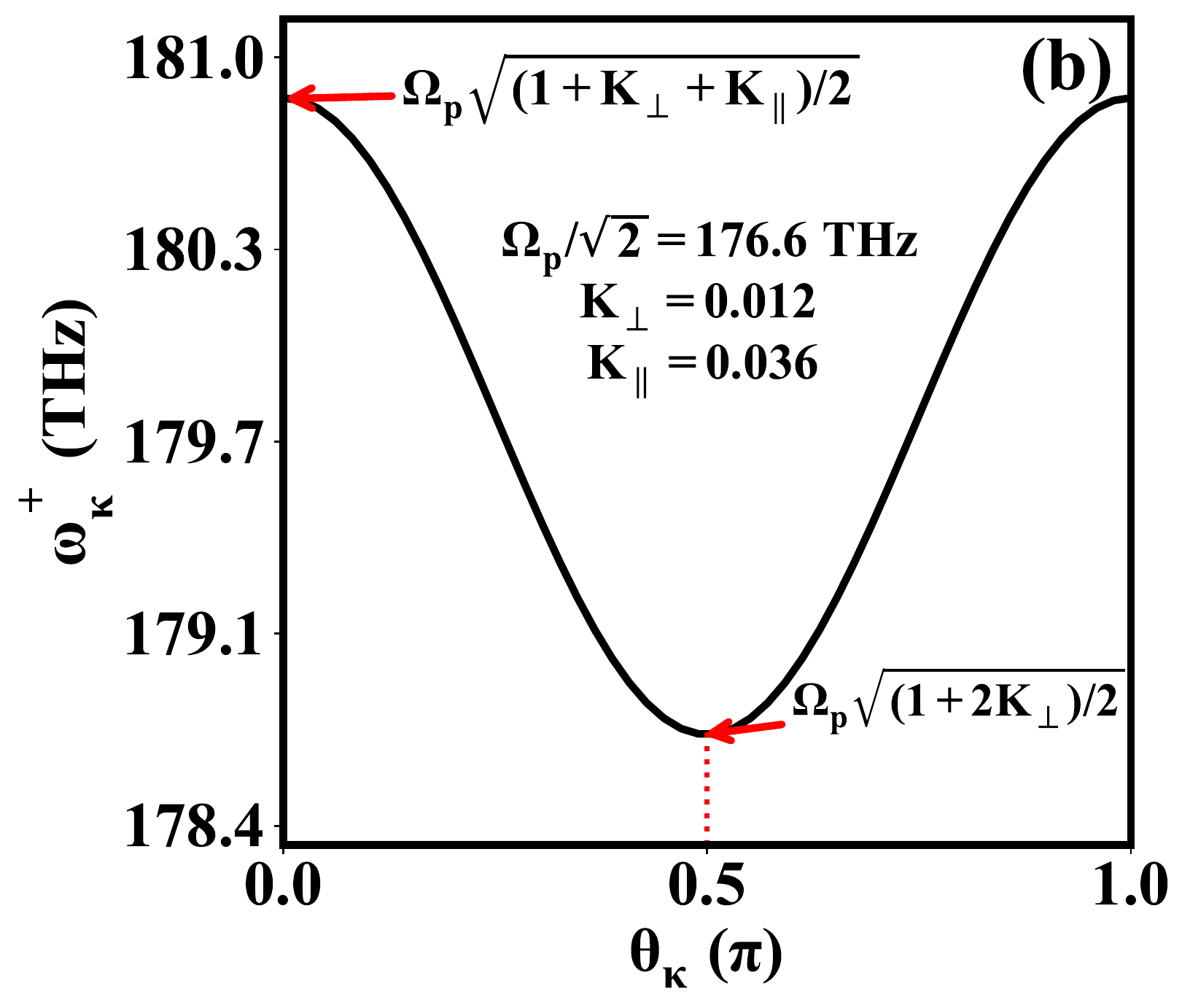}
\hspace{-2.15cm}\includegraphics[width=0.465\linewidth]{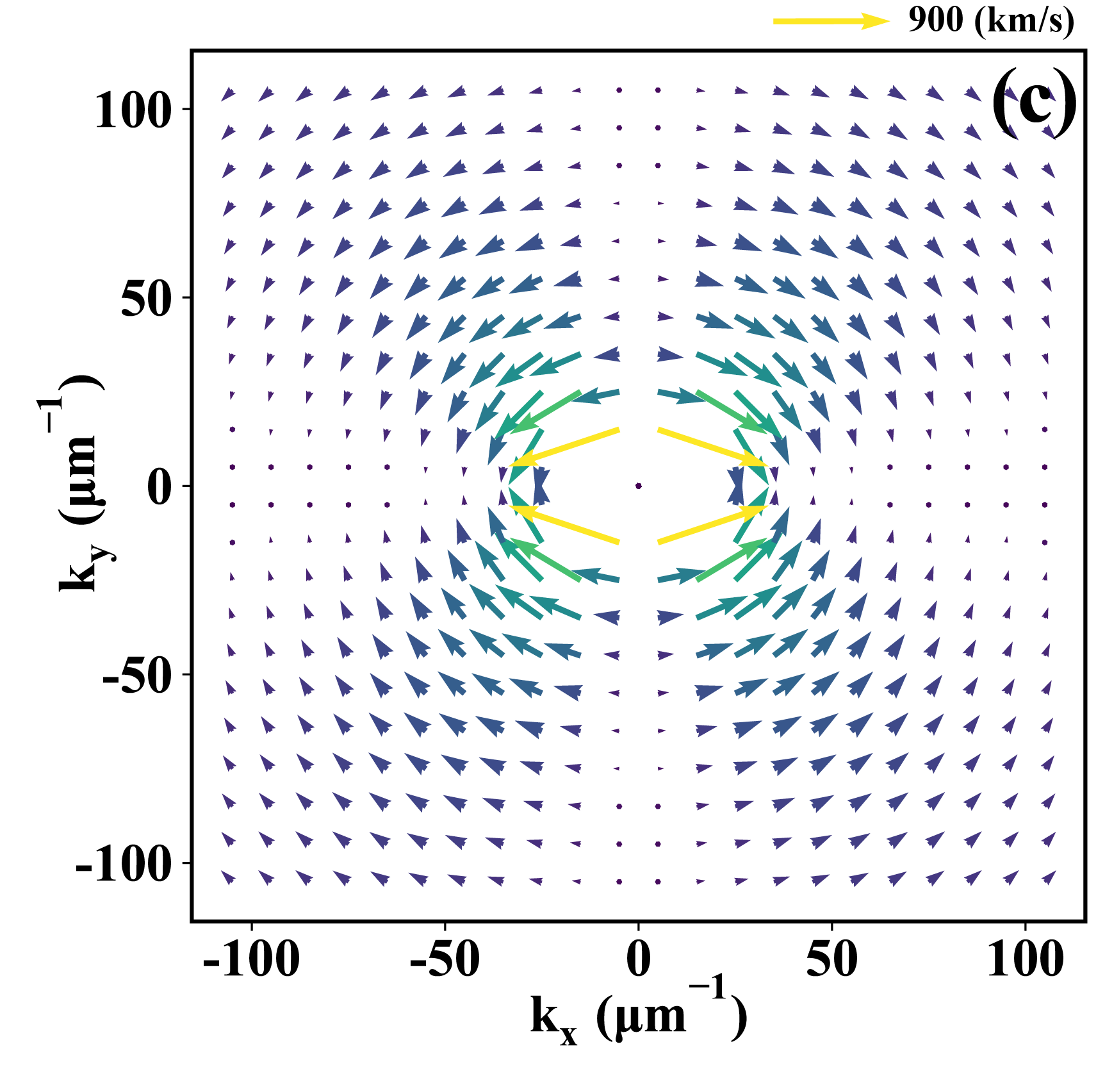}
\hspace{0.15cm}\includegraphics[width=0.505\linewidth]{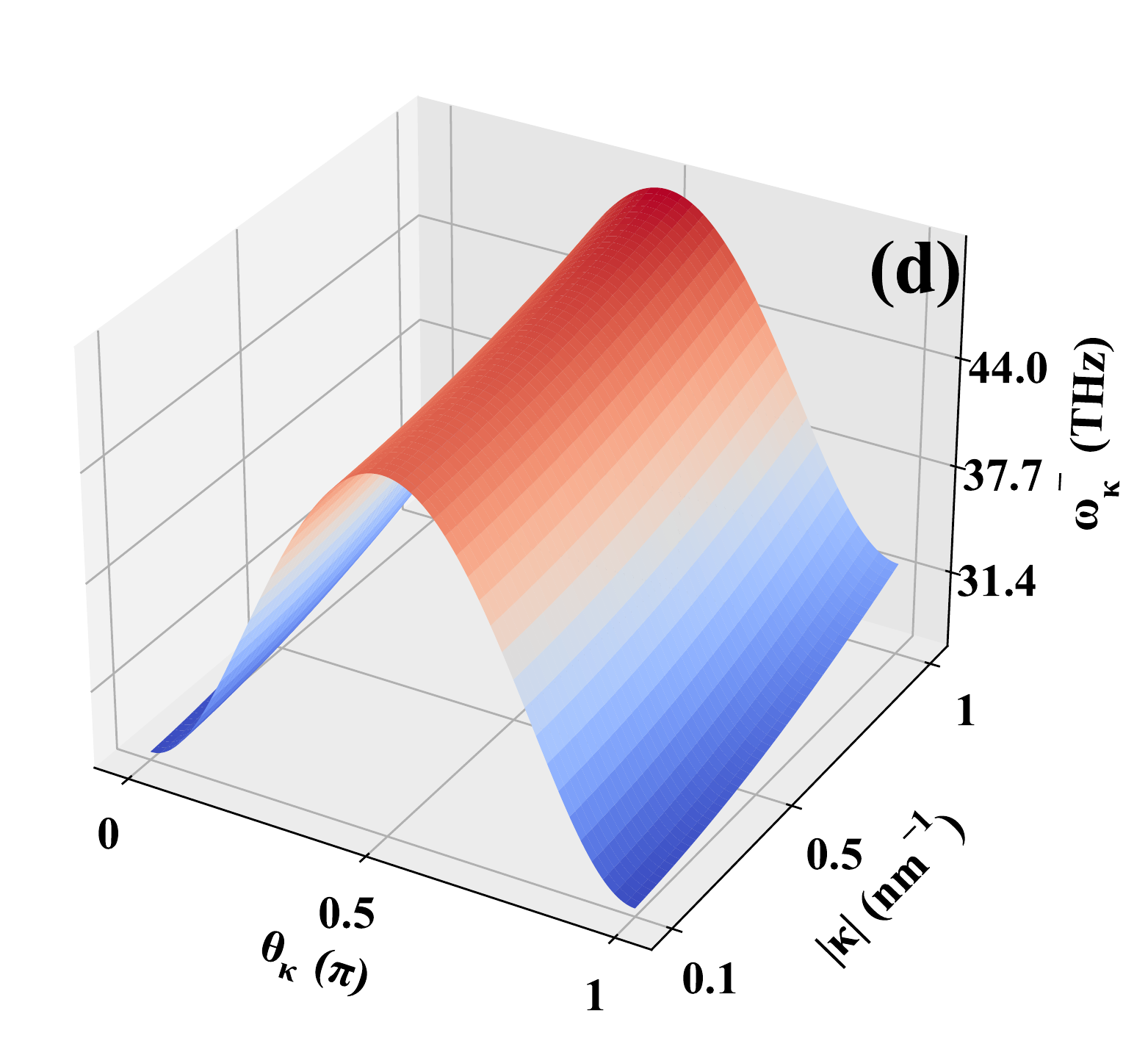}
\caption{Dispersion relation of the surface ferron branches $\omega_{\pmb \kappa}^{-}$ [(a)] and $\omega_{\pmb \kappa}^{+}$ [(b)] for the room-temperature parameters of LiNbO$_3$. At low frequencies, the surface ferrons are strongly anisotropic. They can be excited by focused lasers with propagation directions that can be tuned by its frequency $\omega_0$ (see below). \textcolor{blue}{(c) is a plot of the group velocity vector field of the lower branch, while (d) illustrates the effect of the gradient term of ${\bf p}$ on the dispersion at higher wave numbers.}}
\label{surface_mode_dispersion}
\end{figure}

Figure~\ref{surface_mode_dispersion} illustrates the unique features of the dispersion with the conventional ferroelectric insulator LiNbO$_3$, with which at room temperature the parameters $\alpha=-2.012\times10^9 ~\text{Nm}^2/\text{C}^2$,  $\beta=3.608\times 10^9~\text{Nm}^6/\text{C}^4$, and $\lambda=1.345\times 10^9~\text{Nm}^2/\text{C}^2$ \cite{LiNbO3} lead to $P_0=0.746~{\rm C/m^2}$, $ K_\perp=0.012$, and $K_\parallel=0.036$. The low-frequency branch is located between $ \sqrt{K_{\perp}}\Omega_p  $ when ${\pmb \kappa}\parallel {\bf P}_0$ and $\sqrt{K_{\parallel}}\Omega_p$ when ${\pmb \kappa}\perp {\bf P}_0$.  Decreasing the temperature, the longitudinal stiffness $K_{\parallel}$ is enhanced since $|\alpha|\propto |T-T_c|$, leading to an enhanced peak frequency of low-frequency surface ferron. The high-frequency branch is located within a narrow range between $ \Omega_p\sqrt{(1+2K_{\perp})/2}$ and $\Omega_p\sqrt{(1+K_\perp+K_\parallel)/2}$, exactly around the surface ionic plasma frequency $\Omega_p/\sqrt{2}$.

\textcolor{blue}{The group velocity of surface ferrons does not vanish but is at large wavelengths transverse to the wave-vector direction as shown for the lower branch in Fig.~\ref{surface_mode_dispersion}(c). In the SM \cite{supplement} we show that the gradient term of the electric polarization in the free energy functional ${g}(\nabla {\bf P})^2/2$  substantially enhances the group velocity in the radial direction at larger wave numbers. In Fig.~\ref{surface_mode_dispersion}(d) we plot the dispersion of the lower branch using $g\sim5.39\times10^{-10}~\mathrm{J\cdot m^3/C^2}$ as recommended for LiNbO$_3$ \cite{LiNbO3-1}. }

We next address the characteristics of the surface ferronic modes and their associated electric stray fields, including polarization and chirality. 
We find ${\bf a}=0$ or ${\bf b}=0$ when $k_z=-ic_1\kappa/(1+c_1)$ or $k_z=ic_1\kappa/(1+c_1)$ for the surface modes [Eq.~(\ref{simplified})] \cite{supplement}. Both solutions conclude that the waves are localized near the upper surface of the ferroelectric slab since they
lead to the same exponential decay as a function of distance from the surface according to Eq.~(\ref{plane_wave}). The in-plane momentum ${\pmb \kappa}$ is arbitrary, strikingly different from its magnetic counterpart: surface magnons \cite{Walker_sphere,DE} in ferromagnets are unidirectional at one surface. We list the characteristics of both surface ferrons in the SM \cite{supplement} and focus on the low-frequency branch that allows the directional routing by the frequency of THz light excitation \cite{terahertz}.

The eigenmodes of the low-frequency branch ${\bf p}({\pmb \kappa},z)={\pmb {\cal P}}({\pmb \kappa},z)\exp(i{\pmb\kappa}\cdot {\pmb \rho}-i\omega_{\pmb \kappa}^{-}t)$ and
\begin{align}
	\left( \begin{array}{c}
	{\cal P}_x({\pmb \kappa},z)	\\
	{\cal P}_y({\pmb \kappa},z)	\\
	{\cal P}_z({\pmb \kappa},z)	
\end{array}\right) &=
\xi\left(
\begin{array}{c}
-{\kappa}/({\delta\sin{\theta_{\pmb \kappa}}})	\\      
{\kappa}/({\delta\cos{\theta_{\pmb \kappa}}})	\\      
	i{\kappa}/(1-\delta\sin^2{\theta_{\pmb \kappa}})	 
\end{array}
\right)e^{z/\Delta_{\pmb \kappa}}
\label{eigenmodes}
\end{align} 
strongly depend on the propagation direction. Here $\xi$ is a normalization constant and 
$\Delta_{\pmb \kappa}=(1-\delta\sin^2{\theta_{\pmb\kappa}})/({\kappa\delta\sin^2{\theta_{\pmb\kappa}}})$  is a strongly anisotropic decay length. The polarization of these waves depends on the stiffness difference $\delta =K_\parallel-K_\perp\rightarrow \varepsilon_0(-2\alpha-\lambda)$ that depends on temperature and material. For the conventional materials such as LiNbO$_3$ \cite{LiNbO3} and LiTaO$_3$ \cite{LiTaO_3}, $\delta \ll 1$ at room temperature, such that  $\{|p_x({\pmb \kappa})|,|p_y({\pmb \kappa})|\}\gg |p_z({\pmb \kappa})|$. The fluctuation lies in the plane and, since ${\bf p}({\pmb \kappa})\cdot {\pmb \kappa}=0$, is perpendicular to the momentum. The surface ferrons are therefore  linearly polarized and not chiral.

Ferrons carry polarization and energy in the presence of various driving forces
\cite{ferron_1,ferron_2,ferron_3,ferron_4}. However, surface ferrons also betray their presence by THz evanescent electric stray fields \({\bf E}_{\rm out}^{(d)}(\pmb\kappa ,\pmb \rho , z>0)\) outside the slab. Inserting the eigenmodes (\ref{eigenmodes}) into the Coulomb integral (\ref{integral_G})
\begin{align}
	{\bf E}_{\rm out}^{(d)}({\pmb \kappa})=(\pmb{\hat{\kappa}}+i\hat {\bf z}) (\xi \kappa/\varepsilon_0)e^{-\kappa z}e^{i({\pmb\kappa}\cdot {\pmb \rho}-\omega_{\pmb \kappa}^- t)}. 
	\label{electric_stray_field}
\end{align}
Since divergence free $\nabla\cdot {\bf E}^{(d)}_{\rm out}=0$, this field is circularly polarized with the precession axis or ``spin" governed by the cross product of the momentum propagation direction $ \pmb{\hat{\kappa}}={\pmb\kappa}/\kappa $ and the surface normal $\hat{\bf z}$.

\textit{Directional routing by laser}.---A resonant excitation of the surface ferrons by photons can best be formulated in a quantum language, expanding the fluctuations into surface ferron operators $\hat{a}_{\pmb \kappa}$ and eigenmodes Eq.~(\ref{eigenmodes})  that are normalized as  harmonic oscillators \cite{supplement}: $\hat{\bf p}({\bf r},t)=\sum_{\pmb \kappa}\left({\pmb {\cal P}}({\pmb \kappa},z)\exp(i {\pmb \kappa \cdot\pmb\rho})\hat{a}_{\pmb \kappa }(t)+\text{H.c.}\right)$.
 A laser field ${\bf E}^{(e)}({\bf r})=\sum_{\pmb \kappa }\left({\pmb {\cal E}}({\pmb \kappa },z)e^{i{\pmb\kappa\cdot\pmb\rho}}e^{-i\omega_0t}+\text{H.c.}\right)$ with frequency $\omega_0$ and Fourier components $ {\pmb{\cal E}}(\pmb\kappa,z)$ couples with the surface ferron by the Stark interaction $\hat{H}_c/\hbar=\sum_{\pmb \kappa}\omega_{\pmb \kappa}^{-}\hat{a}_{\pmb \kappa}^{\dagger}\hat{a}_{\pmb \kappa}+\sum_{\pmb \kappa}(g_{\pmb \kappa}e^{i\omega_0t}\hat{a}_{\pmb \kappa}+{\rm H.c.})$, where the coupling constant $g({\pmb\kappa})=(-1/\hbar)\int_{-\infty}^{0}dz {\pmb{\cal  P}}({\pmb \kappa},z)\cdot {\pmb{\cal E}}(\pmb\kappa,z)$ follows the last term in Eq.~(\ref{free_energy}).
In terms of the damping parameter $\gamma$ in Eq.~(\ref{LKTeq}), the ferron decay rate  $\Gamma=\gamma\varepsilon_0  \Omega_p^2/2$.  During the illumination
 \begin{equation}
a_{\pmb \kappa }(t)\equiv \langle \hat{a}_{\pmb \kappa }(t)\rangle=g^*({\pmb \kappa })e^{-i\omega_0t}/(\omega_0-\omega^{-}_{\pmb \kappa }+i\Gamma),
\label{ferron_amplitudes}
\end{equation}
where $\langle\cdots\rangle$ denotes an ensemble average  \cite{supplement}. The polarization propagation depends on the angle with the ferroelectric order according to $\sin^2\theta_{\pmb \kappa}=\left((\omega_0/\Omega_p)^2-K_{\perp}\right)/\delta$ while the momenta are mainly limited by  focus of the exciting electric field.
The spatiotemporal electric polarization in the steady state reads
 \begin{equation}
  {\bf p}({\bf r},t)=\sum_{\pmb \kappa}\frac{\pmb{\cal P}(\pmb{\kappa},z) g^*(\pmb{\kappa})}{\omega_0-\omega_{\pmb\kappa}+i\Gamma}e^{i({\pmb\kappa}\cdot{\pmb \rho}-\omega_0 t)}+\text{H.c.}.
\label{ferron_real_space}   
 \end{equation}

\begin{figure}[h]
	\centering
	\hspace{-0.25cm}
	\includegraphics[width=0.501\linewidth]{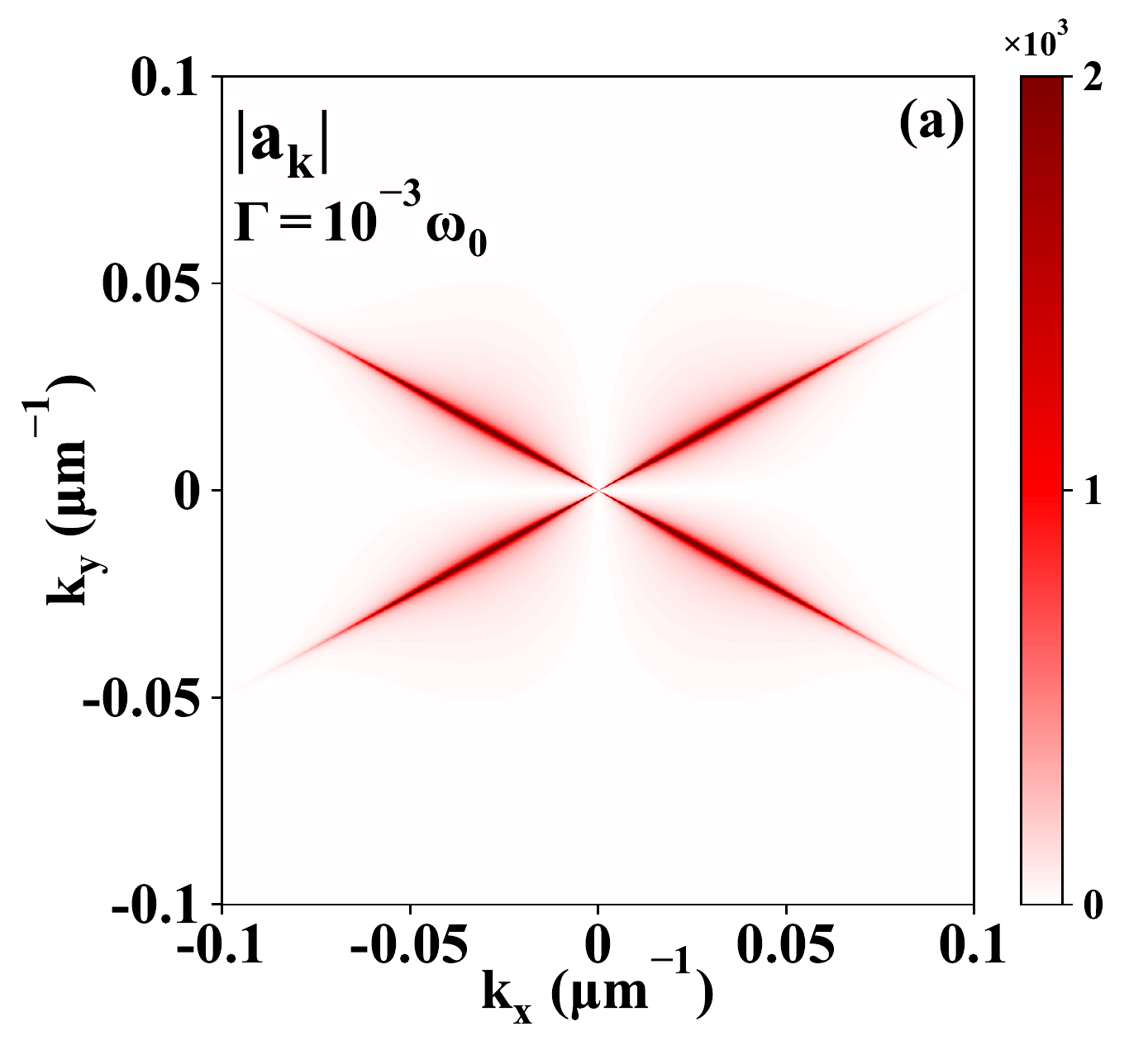}
	\hspace{-0.25cm}
	\includegraphics[width=0.498\linewidth]{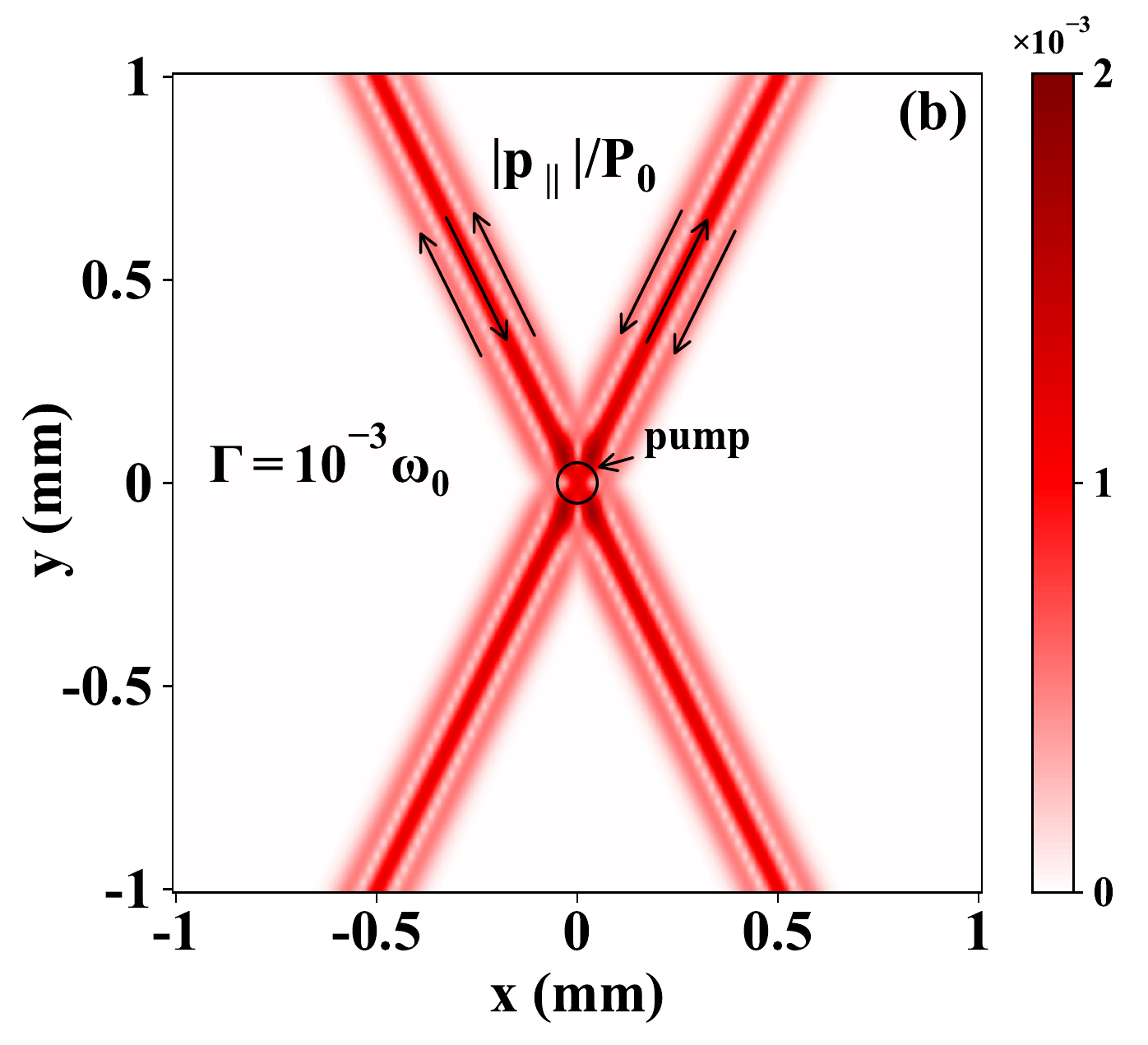}\\
	\hspace{-0.25cm}
	\includegraphics[width=0.495\linewidth]{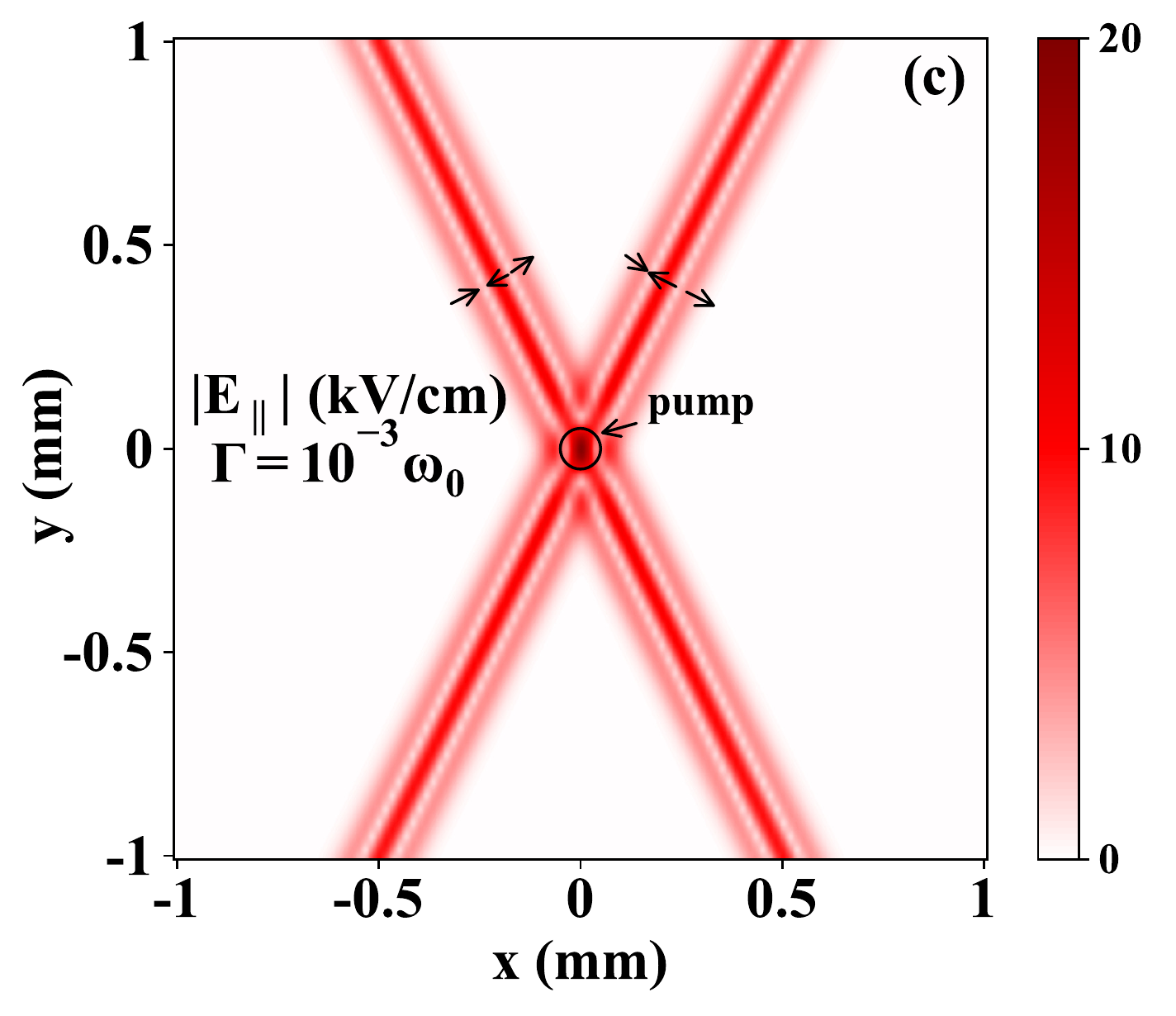}
	\hspace{-0.25cm}
	\includegraphics[width=0.495\linewidth]{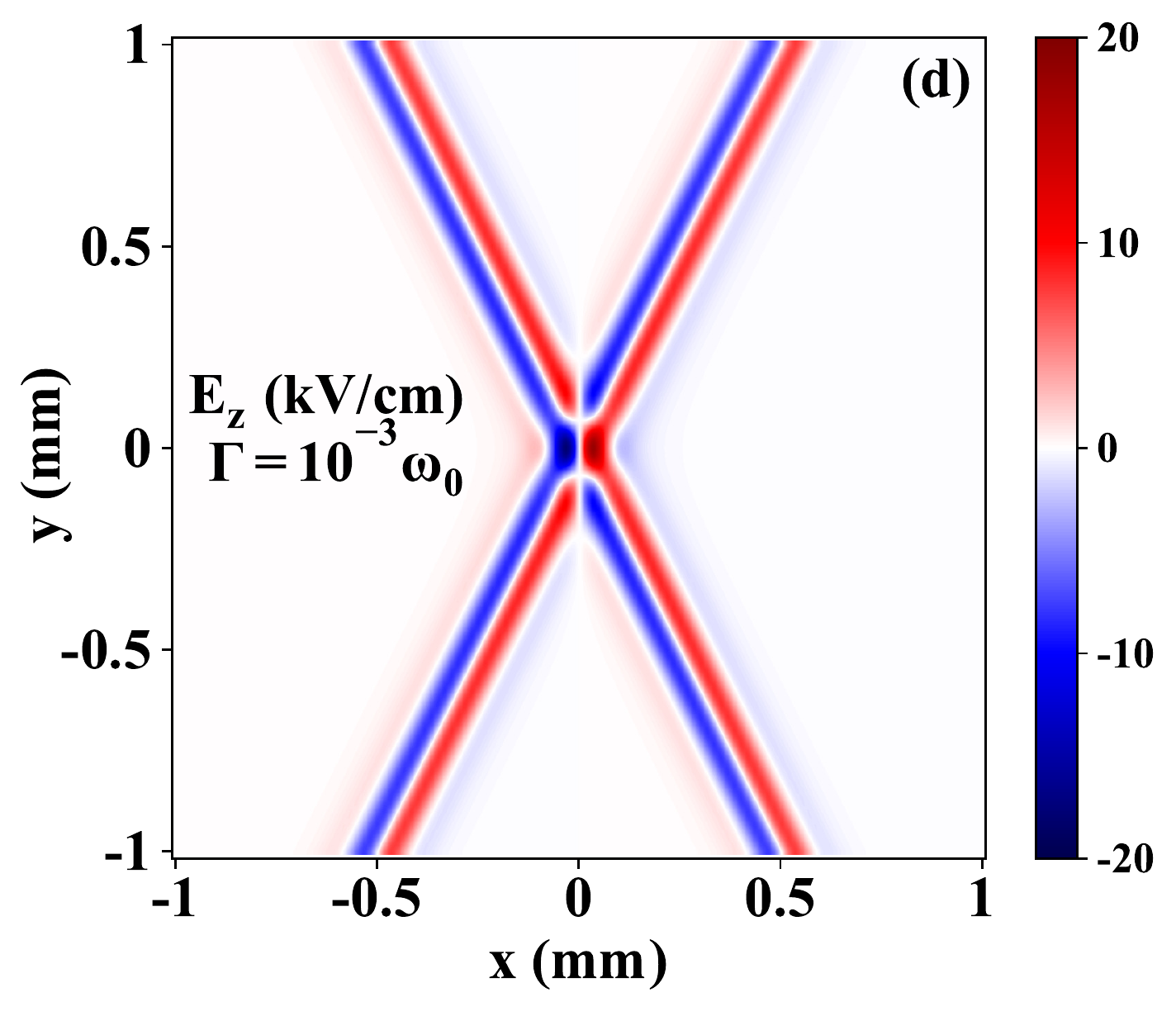}  
	\caption{Directional optical routing of surface ferrons in LiNbO$_3$ resonantly excited by a laser spot.
	(a) addresses the amplitudes of the pumped surface ferrons in reciprocal space. In real space, simultaneously pumped ferrons interfere constructively in micro-scale channels as shown in (b). (c) and (d) plot the in-plane ${\bf E}_{\parallel}$ and out-of-plane ${\bf E}_z$ components of the stray electric field at $z=0$, respectively. The arrows in (b) and (c) are a snapshot of the in-plane electric polarization and the stray electric field.}
\label{ferron_nanorouting}
\end{figure}

  In Fig.~\ref{ferron_nanorouting} we show our predictions of the surface response of a standard ferroelectric such as  the LiNbO$_3$ to a focused THz laser spot with room temperature parameters. We model the laser spot \cite{THz_spot_1,THz_spot_2,THz_spot_3} by a linearly polarized field of Gaussian shape with size $\sigma$, penetration depth $l$, and amplitude $E_0$, i.e., Fourier components ${\pmb {\cal E}}(\pmb\kappa,z)=\pi E_0\sigma^2 \exp[{-(k_x^2+k_y^2)\sigma^2/4}-z^2/l^2] \hat{\bf x} $. We choose parameters $ 2\sigma=100~\mu $m, $ l=200~\mu  $m, $E_0=50$~kV/cm, and $\omega_0=2\pi\times 7$~THz.  A ferron damping rate $\Gamma=10^{-3}\omega_0$ corresponds to a propagation length of up to $\sim 7.7$~mm with an exponential decay  that is much longer than the length scales studied here (refer to SM \cite{supplement} for the fitting decay length).
The surface polarization wave may also be indirectly excited via its coupling to the phonons activated by THz and mid-infrared lasers \cite{MPSD_exp}.  Analogous to the spin wave tomography of magnets \cite{SWAT}, non-resonant broadband excitation by focused optical lasers can reveal the spatiotemporal dynamics of surface ferrons. 

Figure~\ref{ferron_nanorouting}(a) illustrates that the excited surface ferrons propagate preferentially into four directions at an angle $|\theta_0|=|\arcsin\sqrt{[(\omega_0/\Omega_p)^2-K_{\perp}}]/\delta|\sim 63^{\circ}$ with respect to ${\bf P}_0$.   The ``light house" effect, i.e. strong collimation of the surface-ferron beams in real space, as plotted in Fig.~\ref{ferron_nanorouting}(b), is \textcolor{blue}{the caustics caused by the unique group velocity field as plotted in Fig.~\ref{surface_mode_dispersion}(c)},  rendering the characteristic cross-shaped pattern of $|{\bf p}_{\parallel}({\bf r})|=\sqrt{p_x^2({\bf r})+p_y^2({\bf r})}$ at the sample surface $z=0$. The directional routing \cite{nanorouting_1,nanorouting_2} can be tuned \cite{supplement} by varying $\omega_0$. The amplitudes are strongly focused into a tens-of-micrometer channel along $\pi/2-\theta_0$, with respect to ${\bf P}_0$. This direction reflects the Fourier transform of $a({\pmb \kappa})\sim \delta(k_x^2-\tan^2\theta_0k_y^2)$ or $k_x/k_y\sim \tan\theta_0$. The propagation phase in $\exp(i{\pmb \kappa}\cdot {\pmb \rho})\sim \exp[ik_y(y+x\tan\theta_0)]$ after integration over wave vectors favors $y+x\tan\theta_0\sim 0$ or $x/y\sim -\cot\theta_0$ \cite{supplement}. The ac electric polarization flows in the channels both parallel and antiparallel to each other, as indicated by the black arrows in Fig.~\ref{ferron_nanorouting}(b).

The electric polarization produces circularly polarized chiral electric stray fields  Eq.~(\ref{electric_stray_field}) with the same spatial cross-like pattern, as shown in Fig.~\ref{ferron_nanorouting}(c) and (d) for the in-plane ${\bf E}_{\parallel}$ and out-of-plane $E_z$ components at the surface. Measurement of these fields would allow reconstruction of the underlying electric polarization fluctuation. The in-plane component oscillates normal to the channel direction, as illustrated by the black arrows in Fig.~\ref{ferron_nanorouting}(c).  The magnitude of the near electric field can exceed that of the incident laser spot, indicating a substantial plasmonic enhancement \cite{plasmon_1,plasmon_2,plasmon_3,plasmon_4,nanorouting_1,nanorouting_2,polarization_analyzer}.

Equation~(\ref{EOM_fluctuation}) has the same form when the equilibrium polarizations \(\textbf{P}_0 \parallel \hat{\textbf{z}}\) are out-of-plane and lead to surface states with polarization fluctuations normal to the interface with isotropic dispersion. This is again very different from ferromagnets with perpendicular magnetization that support only bulk magnons.

The present calculations have been carried out in an electrostatic approximation, which holds for relatively large wave numbers. For small wave numbers, the ferrons hybridize with the photons or phonons to form surface ferron-photon polaritons or surface ferron-phonon polarons, which we shall address in the future. 
 
\textit{Conclusions.}---In conclusion, we predict that long-range dipolar interaction causes evanescent polarization waves or surface ferrons in uniaxial ferroelectrics. The symmetry breaking by the dipolar order can cause strongly anisotropic  emission patterns from a focused terahertz laser spot. The surface ferrons may be interpreted as gapped Goldstone and Higgs modes of the ferroelectric order. Their  pronounced anisotropy distinguishes them with the surface phonon polariton in polar dielectrics \cite{surface_phonon_polariton_0,surface_phonon_polariton_1,surface_phonon_polariton_2}. These excitations can be monitored by their strong and chiral stray fields.

\vskip0.25cm	
\begin{acknowledgments}
This work is financially supported by the National Natural Science Foundation of China under Grant No.~0214012051, and the startup grant of Huazhong University of Science and Technology (Grants No.~3004012185 and No.~3004012198). P.T. and G.B. acknowledge the financial support by JSPS KAKENHI Grants No.~19H00645 and No.~22H04965. R.S. and S.R. were financially supported in Brazil by Conselho Nacional de Desenvolvimento Cient\'ifico e Tecnol\'ogico (CNPq), Coordenac\~ao de Aperfeicoamento de Pessoal de N\'ivel Superior (CAPES), Financiadora de Estudos e Projetos (FINEP), and Fundac\~ao de Amparo \'a Ci\^encia e Tecnologia do Estado de Pernambuco (FACEPE), and in Chile by Fondo Nacional de Desarrollo Cient\'ifico y Tecnol\'ogico (FONDECYT) Grant No. 1210641.
\end{acknowledgments}


\begin{thebibliography}{99} 

\bibitem{Duality} Duality (electricity and magnetism),  \url{en.wikipedia.org/wiki/Duality_(electricity_and_magnetism)}.

\bibitem{Jackson}  J. D. Jackson, \textit{Classical Electrodynamics}, (Wiley, New York, 1998).

\bibitem{ferron_definition} G. E. W. Bauer, P. Tang, R. Iguchi, and K. Uchida, Magnonics vs. Ferronics, J. Magn. Magn. Mater. \textbf{541}, 168468
(2022).

\bibitem{roadmap} A. Barman \textit{et al.}, The 2021 Magnonics Roadmap, J. Phys.: Condens. Matter, \textbf{33}, 413001 (2021).

\bibitem{PR_insulator} A. Brataas, B. van Wees, O. Klein, G. de Loubens, and
M. Viret, Spin insulatronics, Phys. Rep. \textbf{885}, 1 (2020).

\bibitem{PR_chirality} Yu, Luo, and Bauer, Chirality as Generalized Spin-Orbit Interaction in Spintronics, Phys. Rep. \textbf{1009},1 (2023).

\bibitem{Walker_sphere} L. R. Walker, Magnetostatic Modes in Ferromagnetic Resonance, Phys. Rev. \textbf{105}, 390 (1957).

\bibitem{DE} R. W. Damon and J. R. Eshbach, Magnetostatic modes of a ferromagnet slab, J. Phys. Chem. Solids \textbf{19}, 308 (1961).



\bibitem{heat_conveyer_1} T. An, V. I. Vasyuchka, K. Uchida, A. V. Chumak, K. Yamaguchi, K. Harii, J. Ohe, M. B. Jungfleisch, Y. Kajiwara, H. Adachi, B. Hillebrands, S. Maekawa, and E. Saitoh, Unidirectional spin-wave heat conveyer, Nat. Mater. \textbf{12}, 549 (2013).

\bibitem{heat_conveyer_2} O. Wid, J. Bauer, A. M\"uller, O. Breitenstein, S. S. P. Parkin, and G. Schmidt, Investigation of the unidirectional spin heat conveyer effect in a 200 nm thin Yttrium Iron Garnet film, Sci. Rep. \textbf{6}, 28233 (2016).

\bibitem{heat_conveyer_3} E. Shigematsu, Y. Ando, S. Dushenko, T. Shinjo, and M. Shiraishi, Spin-wave-induced lateral temperature gradient in a YIG thin film/GGG system excited in an ESR cavity, Appl. Phys. Lett. \textbf{112}, 212401 (2018).


\bibitem{heat_conveyer_4} P. Wang, L. F. Zhou, S. W. Jiang, Z. Z. Luan, D. J. Shu, H. F. Ding, and D. Wu, Unidirectional Spin-Wave-Propagation-Induced Seebeck Voltage in a PEDOT:PSS/YIG Bilayer, Phys. Rev. Lett. \textbf{120}, 047201 (2018). 

\bibitem{disorder_1} T. Yu, S. Sharma, Y. M. Blanter, and G. E. W. Bauer, Surface dynamics of rough magnetic films, Phys. Rev. B \textbf{99}, 174402 (2019).

\bibitem{disorder_2}  M. Mohseni, R. Verba, T. Br\"acher, Q. Wang, D. A. Bozhko, B. Hillebrands, and P. Pirro, Backscattering Immunity of Dipole-Exchange Magnetostatic Surface Spin Waves, Phys. Rev. Lett. \textbf{122}, 197201 (2019).

\bibitem{Usami} A. Osada, R. Hisatomi, A. Noguchi, Y. Tabuchi, R. Yamazaki, K. Usami, M. Sadgrove, R. Yalla, M. Nomura, and Y. Nakamura, Cavity Optomagnonics with Spin-Orbit Coupled Photons, Phys. Rev. Lett. \textbf{116}, 223601 (2016).

\bibitem{Sanchar} S. Sharma, Y. M. Blanter, and G. E. W. Bauer, Optical Cooling of Magnons, Phys. Rev. Lett. \textbf{121}, 087205 (2018).

\bibitem{Kei} K. Yamamoto, G. C. Thiang, P. Pirro, K.-W. Kim, K. Everschor-Sitte, and E. Saitoh, Topological Characterization of Classical Waves: The Topological Origin of Magnetostatic Surface Spin Waves, Phys. Rev. Lett. \textbf{122}, 217201 (2019).

\bibitem{surface_ferron} M. G. Cottam, D. R. Tilley, and B. Zeks, Theory of surface modes in ferroelectrics, J. Phys. C: Solid State Phys. \textbf{17}, 1793 (1984).

\bibitem{ferron_1} G. E. W. Bauer, R. Iguchi, and K.-i. Uchida, Theory of Transport in Ferroelectric Capacitors, Phys. Rev. Lett. \textbf{126}, 187603 (2021).

\bibitem{ferron_2} P. Tang, R. Iguchi, K.-i. Uchida, and G. E. W. Bauer, Thermoelectric Polarization Transport in Ferroelectric Ballistic Point Contacts, Phys. Rev. Lett. \textbf{128}, 047601 (2022).

\bibitem{ferron_3} P. Tang, R. Iguchi, K.-i. Uchida, and G. E. W. Bauer, Excitations of the ferroelectric order, Phys. Rev. B \textbf{106}, L081105 (2022).

\bibitem{ferron_4} P. Tang, K.-i. Uchida, and G. E. W. Bauer, Nonlocal Drag Thermoelectricity Generated by Ferroelectric Heterostructures, arXiv:2207.00240.

\bibitem{first_experiment} B. Wooten, R. Iguchi, P. Tang, J. S. Kang, K.-i. Uchida, G. E. W. Bauer, and J. P. Heremans, Sci. Adv. \textbf{9}, eadd7194 (2023).


\bibitem{gibbs} P. Chandra and P. B. Littlewood, A Landau Primer for Ferroelectrics, in \textit{Physics of Ferroelectrics}, (Springer-Verlag, Berlin, Heidelberg, 2007), pp. 69–116.

\bibitem{Weyl} L. Novotny and B. Hecht, \textit{Principles of Nano-Optics}, (Cambridge University Press, Cambridge, England, 2006).


\bibitem{chiral_optics} P. Lodahl, S. Mahmoodian, S. Stobbe, A. Rauschenbeutel, P. Schneeweiss, J. Volz, H. Pichler, and P. Zoller, Chiral quantum optics, Nature (London) \textbf{541}, 473 (2017).

\bibitem{Nori_review} K. Y. Bliokh and F. Nori, Transverse and longitudinal angular momenta of light, Phys. Rep. \textbf{592}, 1 (2015).

\bibitem{Matsuo_review} M. Matsuo, J. Ieda, and S. Maekawa, Mechanical generation of spin current, Front. Phys. \textbf{3}, 54 (2015).

\bibitem {SAW}I. A. Viktorov,  \textit{Rayleigh and Lamb waves: Physical theory and applications} (Plenum Press, New York, 1967).

\bibitem{SOC_light} K. Y. Bliokh, D. Smirnova, and F. Nori, Quantum spin Hall effect of light, Science \textbf{348}, 1448 (2015).


\bibitem{plasmon_1} T. J. Davis and D. E. G\'omez, Colloquium: An algebraic model of localized surface plasmons and their interactions, Rev. Mod. Phys. \textbf{89}, 011003 (2017).

\bibitem{plasmon_2} H. Raether, \textit{Surface Plasmons on Smooth and Rough Surfaces and on Gratings} (Springer, Berlin, 1988).


\bibitem{plasmon_3} S. A. Maier, \textit{Plasmonics: Fundamentals and Applications} (Springer US, New York, 2007).

\bibitem{plasmon_4} J. M. Pitarke, V. M. Silkin, E. V. Chulkov, and P. M. Echenique, Theory of surface plasmons and surface-plasmon polaritons, Rep. Prog. Phys. \textbf{70}, 1 (2006).

\bibitem{nanorouting_1} F. J. Rodr\'iguez-Fortu\~no, G. Marino, P. Ginzburg, D. O'Connor, A. Mart\'inez, G. A. Wurtz, and A. V. Zayats, Near-Field Interference for the Unidirectional Excitation of Electromagnetic Guided Modes, Science \textbf{340}, 328 (2013).

\bibitem{nanorouting_2} J. Petersen, J. Volz, and A. Rauschenbeutel, Chiral nanophotonic waveguide interface based on spin-orbit interaction of light, Science \textbf{346}, 67 (2014).


\bibitem{LKTequation} K. Tani, Dynamics of Displacive-Type Ferroelectrics Soft Modes, J. Phys. Soc. Jpn \textbf{26}, 93 (1969).

\bibitem{LKT2} Y. Ishibashi,  Phenomenological theory of domain walls, Ferroelectrics \textbf{98}, 193 (1989).

\bibitem{LKT3} S. Sivasubramanian, A. Widom, and Y. N. Srivastava, Physical Kinetics of Ferroelectric Hysteresis,
Ferroelectrics \textbf{300}, 43 (2004).

\bibitem{LKT4} A. Widom, S. Sivasubramanian, C. Vittoria, S. Yoon,
and Y. N. Srivastava, Resonance damping in ferromagnets and ferroelectrics, Phys. Rev. B \textbf{81}, 212402 (2010).

\bibitem{GL1} V. A. Zhirnov, A contribution to the theory of domain walls in ferroelectrics, Zh. Eksp. Teor. Fiz. \textbf{35}, 1175 (1958) [Sov. Phys. JETP \textbf{35}, 822 (1959)].

\bibitem{GL2} Y. Ishibashi and E. Salje, A Theory of Ferroelectric 90 Degree Domain Wall, J. Phys. Soc. Jpn. \textbf{71}, 2800 (2002). 

\bibitem{GL3} W. Cao and L. E. Cross, Theory of tetragonal twin structures in ferroelectric perovskites with a first-order phase transition, Phys. Rev. B \textbf{44}, 5 (1991). 
 
\bibitem{GL4} W. Cao, G. R. Barsch, and J. A. Krumhansl, Quasi-one-dimensional solutions for domain walls and their constraints in improper ferroelastics, Phys. Rev. B \textbf{42}, 6396 (1990).

\bibitem{PRB_dipolar} J. Hlinka and P. M\'arton, Phenomenological model of a 90$^{\circ}$ domain wall in BaTiO$_3$-type ferroelectrics, Phys. Rev. B \textbf{74}, 104104 (2006).


\bibitem{PbTiO3} M. J. Haun, E. Furman, S. Jang, H. McKinstry, and L. Cross, Thermodynamic theory of PbTiO3, J. Appl. Phys. \textbf{62}, 3331 (1987).

\bibitem{PbTiO3-1} A. N. Morozovska, E. A. Eliseev, C. M. Scherbakov, and Y. M. Vysochanskii, Influence of elastic strain gradient on the upper limit of flexocoupling strength, spatially modulated phases, and soft phonon dispersion in ferroics, Phys. Rev. B \textbf{94}, 174112 (2016).

\bibitem{LiNbO3} D. A. Scrymgeour, V. Gopalan, A. Itagi, A. Saxena, and P. J. Swart, Phenomenological theory of a single domain wall in uniaxial trigonal ferroelectrics: Lithium niobate and lithium tantalate, Phys. Rev. B \textbf{71}, 184110 (2005).

\bibitem{LiTaO_3} I. Tomeno and S. Matsumura, Dielectric properties of 
LiTaO$_3$, Phys. Rev. B \textbf{38}, 606 (1988).

\bibitem{surface_phonon_polariton_0} R. Fuchs and K. L. Kliewer, Optical Modes of Vibration in an Ionic Crystal Slab, Phys. Rev. \textbf{140}, A2076 (1965).

\bibitem{surface_phonon_polariton_1} H. J. Bakker, S. Hunsche, and H. Kurz, Coherent phonon polaritons as probes of anharmonic phonons in ferroelectrics, Rev. Mod. Phys. \textbf{70}, 523 (1998).

\bibitem{surface_phonon_polariton_2} N. S. Stoyanov, D. W. Ward, T. Feurer, and K. A. Nelson, Terahertz polariton propagation in patterned materials, Nat. Mater. \textbf{1}, 95 (2002).

\bibitem{PRB_Letter} S. H. Zhuang and J.-M. Hu, Role of polarization-photon coupling in ultrafast terahertz excitation of ferroelectrics, Phys. Rev. B \textbf{106}, L140302 (2022).

\bibitem{electronic_ferroelectricity} H. Adachi, N. Ikeda, and E. Saitoh, Ginzburg-Landau action and polarization current in an excitonic insulator model of electronic ferroelectricity, arXiv:2301.11506.

\bibitem{supplement} Supplemental Material [...] for the derivation of the characteristic equations, the surface ferronic modes, the bulk ferronic modes, as well as the quantum approach for the  optical routing of surface ferrons.


\bibitem{Anderson}  P. W. Anderson, Random-Phase Approximation in the Theory of Superconductivity, Phys. Rev. \textbf{112}, 1900 (1958).

\bibitem{Nambu} Y. Nambu, Quasi-Particles and Gauge Invariance in the Theory of Superconductivity, Phys. Rev. \textbf{117}, 648 (1960).

\bibitem{LiNbO3-1} M. S. Richman, X. Li, and A. Caruso, Inadequacy of the extrapolation-length method for modeling the interface of a ferroelectric–graphene heterostructure, J. Appl. Phys. \textbf{125}, 184103 (2019).

\bibitem{terahertz} R. Ulbricht, E. Hendry, J. Shan, T. F. Heinz, and M. Bonn, Carrier dynamics in semiconductors studied with time-resolved terahertz spectroscopy, Rev. Mod. Phys. \textbf{83}, 543 (2011).

\bibitem{THz_spot_1} C. Ruchert, C. Vicario, and C. P. Hauri, Spatiotemporal Focusing Dynamics of Intense Supercontinuum THz Pulses, Phys. Rev. Lett. \textbf{110}, 123902 (2013).

\bibitem{THz_spot_2} X. F. Zang, C. X. Mao, X. G. Guo, G. J. You, H. Yang, L. Chen,
Y. M. Zhu, and S. L. Zhuang, Polarization-controlled terahertz super-focusing, Appl. Phys. Lett. \textbf{113}, 071102 (2018).

\bibitem{THz_spot_3} H. Chen, Z. X. Wu, Z. Y. Li, Z. F. Luo, X. Jiang, Z. Q. Wen, L. G. Zhu, X. Zhou, H. Li, Z. G. Shang, Z. H. Zhang, K. Zhang, G. F. Liang, S. L. Jiang, L. H. Du, and G. Chen, Sub-wavelength tight-focusing of terahertz waves by polarization-independent high-numerical-aperture dielectric metalens, Optics Express \textbf{26}, 29817 (2018).

\bibitem{MPSD_exp} R. Mankowsky, A. von Hoegen, M. F\'orst, and A. Cavalleri, Ultrafast Reversal of the Ferroelectric Polarization, Phys. Rev. Lett. \textbf{118}, 197601 (2017).

\bibitem{SWAT} Y. Hashimoto, S. Daimon, R. Iguchi, Y. Oikawa, K. Shen, K. Sato, D. Bossini, Y. Tabuchi, T. Satoh, B. Hillebrands, G. E. W. Bauer, T. H. Johansen, A. Kirilyuk, T. Rasing, and E. Saitoh, All-optical observation and reconstruction of spin wave dispersion, Nat. Comm. \textbf{8}, 15859 (2017).

\bibitem{polarization_analyzer} A. Espinosa-Soria, F. J. Rodr\'iguez-Fortu\~no, A. Griol, and A. Mart\'inez, On-Chip Optimal Stokes Nanopolarimetry Based on Spin–Orbit Interaction of Light, Nano Lett. \textbf{17}, 3139 (2017).

\end{thebibliography}
\end{document}